# Effects of Demand Variation on Optimal Automated Demand Responsive Feeder Transit System Operation in Rural Areas

Young-Jae Lee, Ph.D.
Associate Professor
Department of Transportation and Urban Infrastructure Studies
Morgan State University
1700 E. Cold Spring Lane, Baltimore, MD 21251
Tel: 443-885-1872; Fax: 443-885-8218; Email: YoungJae.Lee@morgan.edu

Amirreza Nickkar
Ph.D. Student in Transportation and Urban Infrastructure Systems
Morgan State University
Email : amnic1@morgan.edu

Mana Meskar
Ph.D. Student
Department of Industrial Engineering, Sharif University of Technology, Azadi Ave, Tehran, Iran
Tel: +989111553168
Email: mana.meskar@gmail.com

**ABSTRACT**

Improving accessibility is one of the major issues in rural and suburban transportation. With the recent technological improvement of automated vehicles, it is expected that automated demand responsive transit and automated demand responsive feeder transit potentially will be options to improve mobility in rural areas.

One of the main concerns for the optimal automated demand responsive feeder transit operation is variation of passenger demand. Obviously, changes in passenger demand should alter the optimal operation and will result in different passenger travel times and vehicle operation costs.

This paper uses the optimal feeder bus routing algorithm previously developed by the authors. With it, the effects of the various passenger demands with fixed fleet size will be evaluated for the optimal automated demand responsive feeder bus operation based on the example network. The results show that when demand goes up, the maximum average used capacity of vehicles goes up as expected.

Moreover, when demand goes up, transit service becomes more circuitous and makes the passenger costs/person and total costs/person increase, and the service becomes less efficient, although operating costs/passenger decreases.

*Keywords: Feeder Bus, Demand Responsive Transit, Rural Transportation, Automated Transit System, Optimization*

## 1. INTRODUCTION

Fewer mobility options are available in rural areas due to reasons such as distribution of production/attraction centers of trips in a relatively large area and unpredictable travel demand based on low population density in these areas (Velaga et al., 2012). The flexible transit system has been considered as an efficient mobility option in rural areas in many studies; however, the efficiency of these systems is open to discussion because of the different approaches and perspectives toward considering passengers and operator costs (Li & Quadrifoglio, 2010; Mulley & Nelson, 2009).

Previously, the authors developed an algorithm for optimal automated feeder bus routing (Lee et al., 2018; Lee & Nickkar, 2018; Lee et al., 2019),which considers multi-stations and multiple feeder buses while allowing relocations of feeder buses. In this research, using the previously developed algorithm and the model network, the effects of the various passenger demand levels with fixed fleet size will be evaluated for the optimal automated demand-responsive feeder bus operation. The results of this study could be utilized by transportation authorities, transport investment agencies, and collaborators in rural transportation systems.

## 2. LITERATURE REVIEW

The flexible demand-response transport services have been considered both theoretically and practically. Shuttle vans, ring-and-ride services, and dial-up buses are examples of shared demand-response transport services that serve travelers in rural areas. Past studies proved that these systems have the potential to improve mobility efficiency in rural areas not only for travelers but also for travelers with special issues, e.g., the elderly or disabled (Li & Quadrifoglio, 2010; Sloman & Hendy, 2008). Balancing travel demand and service supply to find the desired level of flexibility in mode choice was the main goal of the earliest studies, the majority of which focused on the single vehicle pickup and delivery problem (Psaraftis, 1980; Sexton & Bodin, 1985). Recently, most studies tried to propose more realistic and complicated algorithms by considering multiple passengers and multiple vehicles (Garaix et al., 2010, 2011). A range of attempts to find optimal solution methods have been implemented in past studies: metaheuristics methods (Attanasio et al., 2004; Cordeau & Laporte, 2003), fuzzy logic approaches (Teodorovic & Radivojevic, 2000), integer programming (exact solution) (Cordeau, 2006), and classification methods (Gupta et al., 2007). Although these studies have provided useful results that can minimize passenger or operator costs, there are still limitations in the implementation of these approaches in rural areas. First, most of these studies tried to consider increased operator revenues by scheduling vehicles on optimal routes even though individual passengers' travel time and traveler preferences are important variables that can change the travel behavior of the traveler. Second, these approaches did not consider relocation of fleet service despite the fact that in high-demand conditions fleet relocation might be required. Finally, none of studies considered visualization tools to show optimal solutions. This study tries to address all of these limitations.

## 3. ALGORITHM

The author and co-authors' previously developed algorithm and the model network (Lee et al., 2018) were used to evaluate the impact of the passengers' demand sizes on the feeder bus operation. The algorithm minimizes total cost, including vehicle operating costs and passenger travel time, while individual passengers' maximum travel times are limited within given maximum travel times. Also, this algorithm applies the Simulated Annealing (SA) algorithm to solve the proposed model.

The algorithm starts with finding the shortest distance and the shortest travel time from each passenger to the corresponding station. In order to consider individual passengers' acceptable travel times and acceptable circuity of the routing, additional travel Time Ratio ($R_{ATT}$) and Maximum Additional Travel Time Ratio (Max $R_{ATT}$) as shown in Equations 1 and 2 are introduced in this research. Given Max $R_{ATT}$ and computed shortest travel times are used to define the maximum acceptable travel time for each passenger. Using those maximum acceptable travel times for passengers as constraints, optimal routings are developed for each station using the SA algorithm.

$$\text{Additional Travel Time Ratio } (R_{ATT}) \geq \frac{(Actual\ travel\ time)i}{(Shortest\ travel\ time)\ i} \quad (1)$$

$$\text{Maximum Additional Travel Time Ratio (Max } R_{ATT}) \geq \max \left[\frac{(Actual\ travel\ time)i}{(Shortest\ travel\ time)\ i}\right] \quad (2)$$

$i$ = Individual passenger

The framework of the developed algorithm is shown in Figure 1.

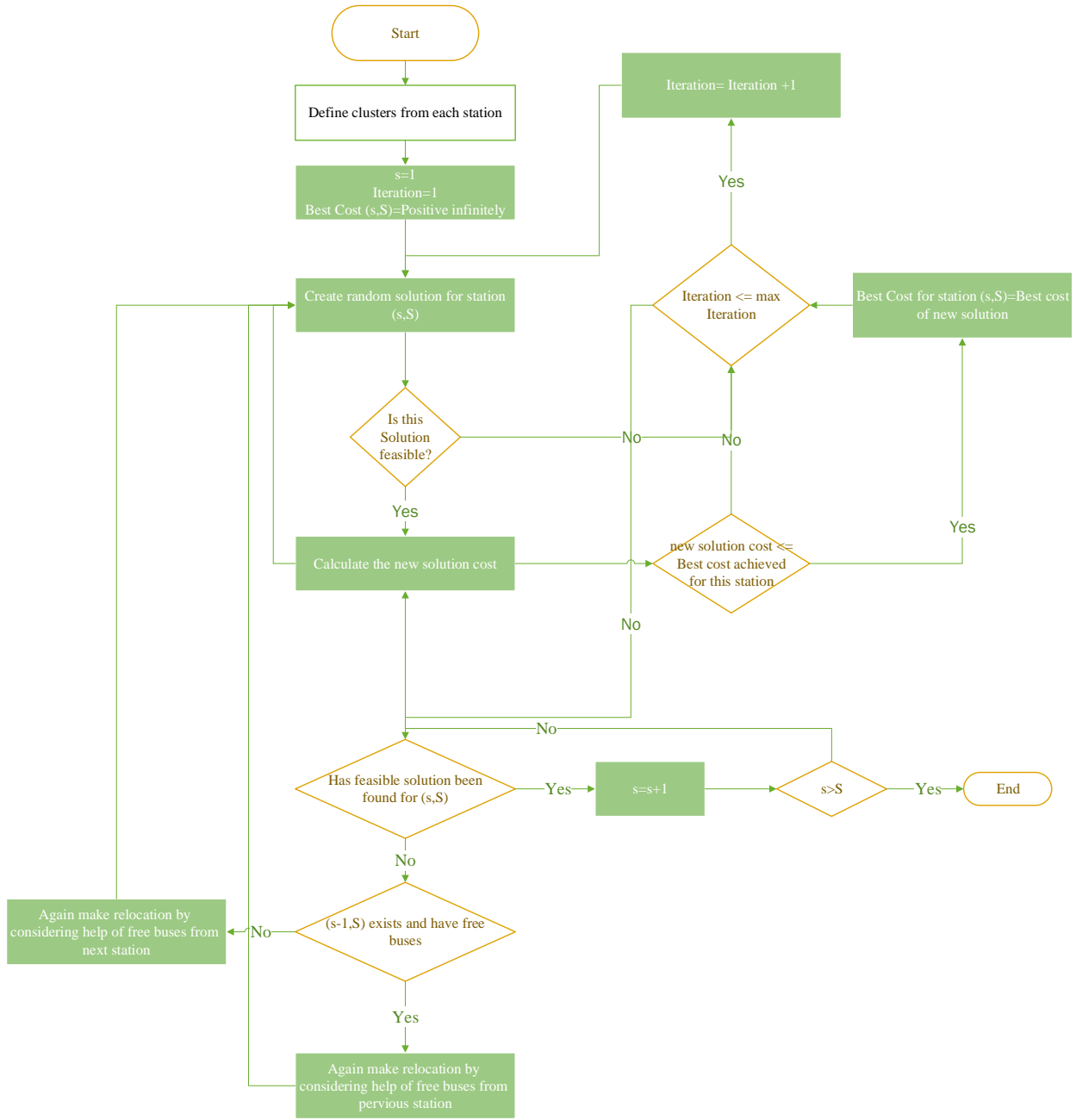

**FIGURE 1 The conceptual flowchart for the proposed algorithm to solve the problem (Lee et al., 2018).**

A hypothetical rail transit line that has four stations fitted to rural conditions was considered to examine and evaluate the efficiency of the algorithm. In this example, the headway of the train is assumed to be 20 minutes and the travel time between two stations is assumed to be two minutes. The capacity for each bus is assumed to be eight passengers. Passengers' boarding and alighting time at the nodes and the stations have been waived in this study. The conceptual operation of the algorithm is shown in Figure 2. Table 1 shows the number of boarding and alighting (B/L) passengers for each station and each train. For example, in Station 1 and for Train 1, 10 passengers need to be picked up and get on, and 20 passengers get off and need to be at Station 1 for Train 1. In this paper, it is assumed that the average speed for feeder buses is 30 km/h and for trains is 60 km/h; each bus has a 15-passenger capacity, and the distance between stations is 4 km. The travel time monetary value for each passenger has been placed at $20 per hour, and $5 per kilometers for vehicles is used as the travel cost.

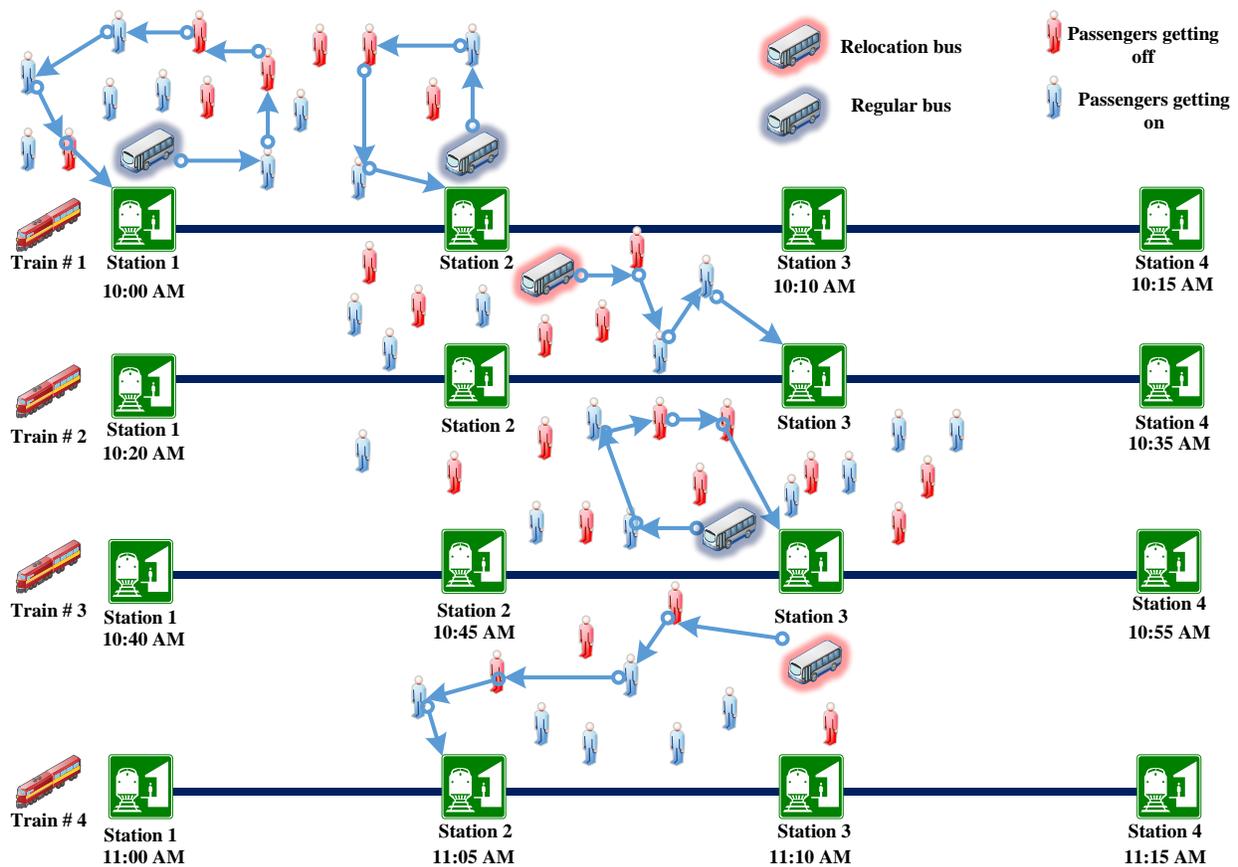

**FIGURE 2 Conceptual operation of feeder transit (regular and relocation buses) (Lee et al., 2018)**

**TABLE 1 Passenger Information for Each Station and Each Train**

| Scenario | Passengers status in stations | Station 1 | | Station 2 | | Station 3 | | Station 4 | | Average total direct distance (Km) |
|---|---|---|---|---|---|---|---|---|---|---|
| | | (B/L) | Average direct distance (Km) | (B/L) | Average direct distance (Km) | (B/L) | Average direct distance (Km) | (B/L) | Average direct distance (Km) | |
| 1 (Total demand = 120) | Train 1 | 10+20 | 1.66 | 17+13 | 1.30 | 13+17 | 1.52 | 16+14 | 1.57 | 1.51 |
| | Train 2 | 8+12 | 1.49 | 15+25 | 1.51 | 15+20 | 1.41 | 10+15 | 1.52 | 1.48 |
| | Train 3 | 20+20 | 1.47 | 13+7 | 1.72 | 7+8 | 1.54 | 24+21 | 1.69 | 1.61 |
| | Train 4 | 18+12 | 1.59 | 14+16 | 1.71 | 12+18 | 1.66 | 13+17 | 1.55 | 1.63 |
| 2 (Total demand = 140) | Train 1 | 14+21 | 1.46 | 17+18 | 1.44 | 18+17 | 1.57 | 18+17 | 1.43 | 1.47 |
| | Train 2 | 12+13 | 1.61 | 23+22 | 1.56 | 22+18 | 1.62 | 11+19 | 1.52 | 1.58 |
| | Train 3 | 24+21 | 1.48 | 12+13 | 1.39 | 7+13 | 1.47 | 23+27 | 1.51 | 1.46 |
| | Train 4 | 19+16 | 1.59 | 18+17 | 1.58 | 15+20 | 1.62 | 19+16 | 1.50 | 1.57 |
| 3 (Total demand = 160) | Train 1 | 17+23 | 1.60 | 19+21 | 1.52 | 21+19 | 1.53 | 18+22 | 1.45 | 1.53 |
| | Train 2 | 16+14 | 1.35 | 25+25 | 1.44 | 12+8 | 1.63 | 29+31 | 1.58 | 1.50 |
| | Train 3 | 5+5 | 1.54 | 37+33 | 1.52 | 29+11 | 1.51 | 5+15 | 1.48 | 1.51 |
| | Train 4 | 17+23 | 1.53 | 22+18 | 1.64 | 17+23 | 1.55 | 18+22 | 1.42 | 1.54 |

## 4. ANALYSES AND RESULTS

The input variables in the algorithm are: passenger demand coordinations 20 minutes before train arrival, vehicle speed, trains' schedules, stations' coordination, and velocity of trains. In the next step, the algorithm finds the optimal solution. It is important in this algorithm that the cost calculation process includes three parameters: without help, with help from the previous station, and help from the next station. Finally, the outputs would be passenger travel times, vehicle traveled distances, assigned buses in each station in each time window, relocated buses, and routes.

Figure 3 shows the results of the feeder bus routings including relocation of the buses. For Train 2, one bus was relocated, shown in dash lines. The algorithm calculated each individual passenger's shortest direct travel time from the origin to the station (or to the destination from the station), and computed the maximum acceptable travel time in the feeder bus as a constraint for the algorithm. Those acceptable additional times are calculated and used as a ratio (travel time in the feeder bus/direct travel time to the origin or to the station). Then, three different scenarios by different passengers' demand levels and Max $R_{ATT}$ of 3 were applied to the algorithm. In the first scenario it assumed that the total demand is 120 passengers, for the second scenario 140 passengers, and 160 passengers for the third scenario.

Tables 2 and 3 summarize the results of the comparison of models with different demands. The maximum used capacity of vehicle (MUCV) for the first and third scenarios is 12 seats while for the second scenario it is 11. The maximum average used capacity of vehicle (MAUCV) for demand of the first scenario is equal to 2.31, which differs from the second scenario (MUCV=2.71) and the third scenario (MUCV=2.88). In transit operation, usually, if demand goes up, the transit system becomes more efficient (less passenger travel time/person and total costs/person) by making more direct services by increased fleet size. However, in this paper, the results show that when the fleet size is fixed, transit service becomes more circuitous and makes the passenger costs/person and total costs/person go up, and the service becomes less efficient when demand goes up, although it was also found that operating costs/passenger decreases.

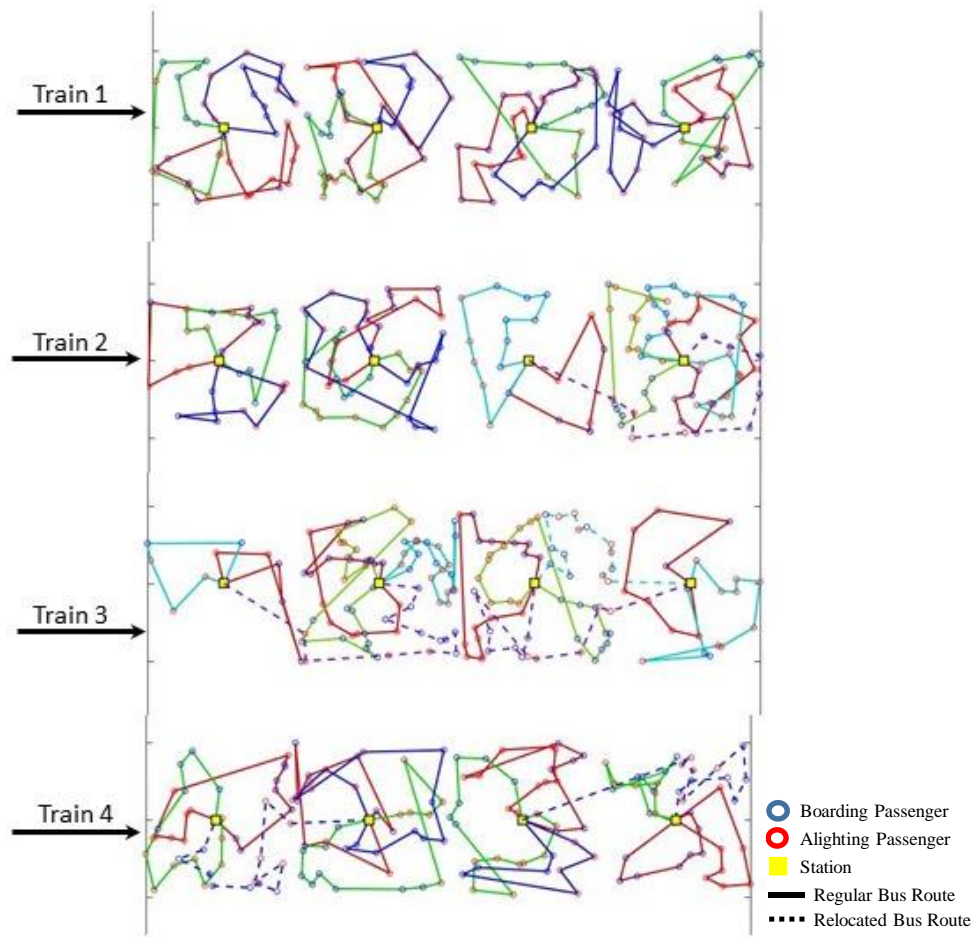

**FIGURE 3 Results of the feeder bus routings (Scenario 3)**

**Table 2 Summary of the Routings for Various Scenarios**

| Models | Passenger related factors | | | | | | | Agency related factors | | | | Total | |
|---|---|---|---|---|---|---|---|---|---|---|---|---|---|
| | Total passenger average direct travel distance (Km) | Total passenger average distance traveled (Km) | Circuity of passenger travels due to feeder bus routings | Average travel time per passenger (h) | Average passengers total travel time (h) | Average travel costs per passenger ($) | Total passenger travel cost ($) | Total bus trips | Total vehicles traveled distance (Km) | Total vehicle operating cost ($) | Total vehicle operating costs/ total passengers ($) | Total Cost ($) | Total cost per passenger ($) |
| Scenario 1 | 1.55 | 2.75 | 1.76 | 0.79 | 2.44 | 1.63 | 781.12 | 48 | 432.64 | 129.79 | 0.270 | 910.91 | 1.90 |
| Scenario 2 | 1.52 | 3.37 | 2.21 | 0.85 | 2.99 | 1.71 | 958.55 | 48 | 498.86 | 149.66 | 0.267 | 1108.21 | 1.98 |
| Scenario 3 | 1.51 | 4.03 | 2.65 | 0.88 | 3.58 | 1.79 | 1145.76 | 48 | 532.64 | 159.79 | 0.250 | 1305.55 | 2.04 |

**Table 3 Variation of MAUV and MAUCV in Different Scenarios**

| Train | Stations | MUCV | | | MAUCV | | |
|---|---|---|---|---|---|---|---|
| | | Demand 120 | Demand 140 | Demand 160 | Demand 120 | Demand 140 | Demand 160 |
| Train 1 | station 1 | 8 | 7 | 8 | 2.67 | 2.33 | 2.67 |
| | station 2 | 7 | 8 | 11 | 2.33 | 2.67 | 3.67 |
| | station 3 | 6 | 8 | 9 | 2.00 | 2.67 | 3.00 |
| | station 4 | 6 | 8 | 8 | 2.00 | 2.67 | 2.67 |
| Train 2 | station 1 | 5 | 6 | 7 | 1.67 | 2.00 | 2.33 |
| | station 2 | 10 | 9 | 11 | 3.33 | 3.00 | 3.67 |
| | station 3 | 9 | 10 | 7 | 3.00 | 3.33 | 3.50 |
| | station 4 | 7 | 9 | 10 | 2.33 | 3.00 | 2.50 |
| Train 3 | station 1 | 8 | 10 | 3 | 2.67 | 3.33 | 1.50 |
| | station 2 | 6 | 5 | 12 | 2.00 | 1.67 | 3.00 |
| | station 3 | 3 | 9 | 10 | 1.00 | 4.50 | 2.50 |
| | station 4 | 9 | 9 | 6 | 3.00 | 2.25 | 3.00 |
| Train 4 | station 1 | 6 | 7 | 8 | 2.00 | 2.33 | 2.67 |
| | station 2 | 6 | 7 | 9 | 2.00 | 2.33 | 3.00 |
| | station 3 | 8 | 9 | 11 | 2.67 | 3.00 | 3.67 |
| | station 4 | 7 | 9 | 8 | 2.33 | 3.00 | 2.67 |

## 5. CONCLUSIONS

In this paper, sensitivity analysis for various demand sizes has been performed to evaluate the performance of an optimal feeder bus routing algorithm developed by the authors. (Lee et al., 2018). In transit operation, usually, if demand goes up, the transit system becomes more efficient (less passenger travel time/person and total costs/person) by making more direct services by increased fleet size. However, in this paper, the results show that when the fleet size is fixed, transit service becomes more circuitous and makes the passenger costs/person and total costs/person go up, and the service becomes less efficient when demand size goes up, although it was also found that operating costs/passenger decreases. In the future, the relationship between demand size, vehicle capacity, fleet size, costs/person and service efficiency will be further examined with demand changes.

## 6. ACKNOWLEDGMENT

This research was supported by the Urban Mobility & Equity Center at Morgan State University and the University Transportation Center Program of the U.S. Department of Transportation.